\documentclass[print]{aa}  

\usepackage{multirow}
\usepackage{amsmath}
\usepackage[colorlinks,linkcolor=blue,anchorcolor=blue,citecolor=blue]{hyperref}
\usepackage{graphicx,subfig} 
\usepackage{lscape}
\usepackage{longtable}
\usepackage{txfonts}
\usepackage{natbib}
\usepackage{color}
\usepackage{booktabs}

\usepackage{mathrsfs}

\title{Quasi-periodic Eruptions from Helium Envelope of Hydrogen-deficient Stars Stripped by Supermassive Black Holes}

\author{Z. Y. Zhao\inst{1},
	Y. Y. Wang\inst{2},
	Y. C. Zou\inst{3},
	F. Y. Wang\inst{1,4}\thanks{e-mail: \href{fayinwang@nju.edu.cn}{\tt fayinwang@nju.edu.cn}},
	Z. G. Dai \inst{5,1},
}

\institute{
	\inst{1}{School of Astronomy and Space Science, Nanjing University, Nanjing 210093, China}\\
	\inst{2}{Anton Pannekoek Institute for Astronomy, University of Amsterdam, Science Park 904, 1098 XH Amsterdam, The Netherlands}\\
	\inst{3}{School of Physics, Huazhong University of Science and Technology, Wuhan 430074, China}\\
	\inst{4}{Key Laboratory of Modern Astronomy and Astrophysics (Nanjing University), Ministry of Education, Nanjing 210093, China}\\
	\inst{5}{Department of Astronomy, School of Physical Sciences, University of Science and Technology of China, Hefei 230026, Anhui, China}\\
}

\titlerunning{QPEs from He envelopes Stripped by SMBHs}
\authorrunning{Zhao et al.}

\begin{document}

\abstract
{Quasi-periodic eruptions (QPEs), which are a new kind of X-ray bursts with a recurrence time of several hours, have been detected from supermassive black holes (SMBHs) in galactic nuclei. Recently, the two QPEs discovered by the \textit{eROSITA} show asymmetric light curves with a fast rise and a slow decline. Current models cannot explain the observational characteristics of QPEs. Here we show that QPEs can be generated from the Roche lobe overflows at each periapsis passage of an evolved star orbiting an SMBH. The properties of the companion stars are constrained via analytic estimations. We find that hydrogen-deficient post-AGB stars are promising candidates. Modules for Experiments in Stellar Astrophysics (MESA) stellar evolution code is used to construct the hydrogen-deficient stars which can fulfill the requirements, as obtained through analytical estimates, to produce the properties of QPEs, including the fast-rise and slow-decay light curves, periods, energetics, and rates. Furthermore, the extreme mass ratio $\sim 10^5$ between the SMBH and the donor will lead to a phenomenon called extreme mass-ratio inspiral (EMRI), producing millihertz gravitational waves. These QPEs would be detected as EMRI sources with electromagnetic counterparts for space-based GW detectors, such as Laser Interferometer Space Antenna (LISA) and Tianqin. They would provide a new way to measure the Hubble constant and further test the Hubble constant tension.}


\maketitle

\section{Introduction}\label{intro}
Quasi-periodic eruptions (QPEs) are a new kind of sudden X-ray brightening from the region near the supermassive black holes (SMBHs) of galactic nuclei, both active \citep{Miniutti_2019,Giustini_2020} and quiescent \citep{Arcodia_2021}. The peak luminosities of the X-ray bursts are $\sim 10^{41}-10^{43}$ erg in 0.5–2 keV energy band. The average recurrence time of QPEs is $\sim 2.4-18$ h with the duty cycle $\sim 6\% - 41\%$. No optical/UV counterpart has been reported for them. 

The first QPE was discovered with \textit{XMM-Newton} in the central region of Seyfert 2 galaxy GSN 069 \citep{Miniutti_2019}. The decay of X-ray emission of GSN 069 could be a tidal disruption event (TDE) source according to the long-term evolution of X-ray flux and spectrum \citep{Shu_2018} and the carbon and nitrogen ratio [C/N] abundance \citep{Sheng_2021}. A second QPE system was discovered in RX J1301.9+2747 \citep{Giustini_2020}. More recently, two more QPEs, eRO-QPE1 and eRO-QPE2, were discovered
with \textit{eROSITA} instrument on the Spectrum-Roentgen-Gamma (SRG) space observatory \citep{Arcodia_2021}. Follow-up observations by the \textit{XMM-Newton} X-ray
telescope and \textit{NICER} confirmed the bursting nature of the two sources. The eRO-QPE1 and eRO-QPE2 expand QPE durations and recurrence times towards longer and shorter timescales, respectively. 
The light curves of them are asymmetric, with a fast rise and a slow decay \citep{Arcodia_2021}. 
The first QPE  in GSN 069 also shows asymmetric behavior from its phase-folded light curve.

The SMBHs associated QPEs have typical low masses. For GSN 069, the mass of SMBH is $4\times 10^5~M_\odot$ \citep{Miniutti_2019}. 
From the X-ray and UV analysis, the centeral SMBH mass of RXJ1301.9+2747 is found to be $0.8-2.8 \times 10^6~M_\odot$ \citep{Shu_2017}. 
The total stellar masses are estimated to be $3.8^{+0.4}_{-1.9}\times10^9~M_\odot$ and $1.01^{+0.01}_{-0.5}\times10^9~M_\odot$ from the optical spectra
for eRO-QPE1 and eRO-QPE2 \citep{Arcodia_2021}, indicating the masses of central SMBHs are in the range of $10^5-10^6~M_\odot$ from the scaling relation between central SMBH mass and total galaxy stellar mass \citep{Reines_2015}. The properties of the four QPEs are listed in Table \ref{QPEs}.

\begin{table*}
	\centering
	\caption{The observed properties of quasi-periodic eruptions events.}
	\label{QPEs}
	\begin{tabular}{cccccc}
		\hline
		\hline
		Source & Redshift &Duration (h) & Recurrence Time (h) & $L_{\text{p}}$ (erg s$^{-1}$) & References\\
		\hline
		GSN 069 &0.018 & 0.54   & 9   & 5 $\times$ 10$^{42}$ & \cite{Miniutti_2019} \\
		RXJ1301.9+2747 &0.02358 &  $\sim 0.3$    &   $\sim 3-5$   &               1.5 $\times$ 10$^{41}$ & 
		\cite{Giustini_2020} \\ 
		eRO-QPE1 &0.0505    & 7.6 & 18.5  & 2 $\times$10$^{43}$ & \cite{Arcodia_2021}  \\
		eRO-QPE2 &0.0175   & 0.45 & 2.4 & 10$^{42}$ & \cite{Arcodia_2021}  \\
		\hline
		\hline
	\end{tabular}
\end{table*}

There have been some theoretical models to explain the unusual characteristics of QPEs. For the first QPE, the radiation-
pressure accretion disk instability \citep{Janiuk_2002,Merloni_2006,Janiuk_2011,Grzedzielski_2017} was proposed to interpret it \citep{Miniutti_2019}. However, the quiescent galactic nuclei origin of eRO-QPE1 and eRO-QPE2, together with the fast-rise and slow-decay light curves, would challenge this explanation because the stable rise and rapid decay light curves are predicted \citep{Czerny_2009,Wu_2016}. The gravitational self-lensing binary SMBH model can generate the flare spacings \citep{Ingram_2021}, but cannot reproduce the observed flare profile. Periodic variability is also often associated with a binary orbital period. If the orbital evolution is dominated by gravitational wave emission, the observed properties of eRO-QPE1 and eRO-QPE2 require the mass of the companion to be much smaller than that of the main body \citep{Arcodia_2021}, called extreme mass-ratio inspirals (EMRIs). A low-mass white dwarf (WD) EMRI on a highly eccentric orbit has been proposed to explain the QPEs in GSN 069 \citep{King_2020}, which was first proposed by \citet{Zalamea_2010}. However, we find that the model of WD companion cannot explain other three QPEs (RXJ1301.9+2747, eRO-QPE1 and eRO-QPE2, see Section \ref{model}). More recently, a complex model with a large degree of uncertainty (both observational
and theoretical) was built, in which the QPEs are due to periodic close interactions between two coplanar stellar
EMRIs \citep{Metzger_2021}. 

The fast rise and slow decay of the light curves, combing the repetition property and the spatial consistency with SMBHs, imply that the profile of QPEs resembles that of a tidal stripping event \citep{Shen_2019}. Although the nature of the rise and decay of the light curves is complex, the dynamical timescale of the companion is the key parameter. The more compact the objects, the faster the flux has been found \citep{Guillochon2013,Law-Smith_2017}. For CO WDs, the rise time of flux is about one minute, and for sun-like stars or red giants, the rise time of flux is a few days. The rise time of QPEs is $\sim 0.2-3$ h, indicating that they cannot produce QPEs. We find that hydrogen-deficient post-Asymptotic Giant Branch (post-AGB) stars which remove H envelopes after very late thermal pulse (VLTP) phase \citep{Herwig1999} can satisfy the conditions for QPEs.

Our model is shown as following. For a low-or-intermediate-mass star  (initial mass $1M_\odot<M<10 M_\odot$) after the MS, it will evolve onto the red giant, AGB, post-AGB, the central star of planetary nebulae (CSPNe), and finally WD successively. It is captured by the SMBH at the post-AGB stage, leading to an eccentric binary. It can undergo Roche lobe overflow (RLOF) at periapsis, resulting QPEs. Our model is built on \cite{Wang_2019}, who studied the electromagnetic counterpart of EMRIs produced by the tidal stripping of the He envelope of a massive star by an SMBH.  Motivated by eruption recurrence of cataclysmic variable binaries (CV, e.g., V803 Centauri, Patterson et al. \citeyear{Patterson_2000}), we consider the mass transfer is driven by angular momentum loss caused by gravitational radiation \citep{Paczynski_1981}. This scenario can quantitatively account for the fast rise and slow decay light curves, periods, energetics, and rates of the QPE phenomenon. Moreover, these four QPE systems are detectable sources
of gravitational waves (GW) in the Laser Interferometer Space Antenna (LISA; Danzmann et al. \citeyear{Danzmann_2000}; Amaro-Seoane
et al. \citeyear{Amaro-Seoane_2017}; Amaro-Seoane \citeyear{Amaro-Seoane_2018}) and Tianqin \citep{Luo_2016}.

This paper is organized as follows. The constraints on the properties of the companion star are given in Section \ref{model}. Modules for Experiments in Stellar Astrophysics (MESA) stellar evolution code is used to construct the hydrogen-deficient post-AGB stars, and the results are shown in Section \ref{stars}. We estimate the event rate of QPEs in Section \ref{event_rate}. A discussion of QPEs as the promising GW sources for LISA and Tianqin is given in Section \ref{gw} and a summary will be shown in Section \ref{conclusion}.

\section{Constraints on the properties of the companion star}\label{model}
The tidal interaction between the SMBH and the companion can be described as the penetration factor $\beta= R_{\text{T}}/R_{\text{p}}$, where $R_{\text{T}}$ is the tidal radius and $R_{\text{p}}$ is periapsis. For $\beta \gtrsim 1$, the star will be fully disrupted. The conditions for full tidal disruption can be found in \cite{Rosswog_2009} and \cite{Gezari_2021}. For $0.6 \lesssim \beta<1$, partial disruptions will occur \citep{Guillochon2013}. If $\beta<0.6$, the mass loss of the companion is slow, which is so called tidal disruption near miss \citep{King_2020}. The onset of mass-loss can be assumed when the star just fills the Roche lobe at periapsis.

\begin{figure}
	\includegraphics[width = 0.5\textwidth]{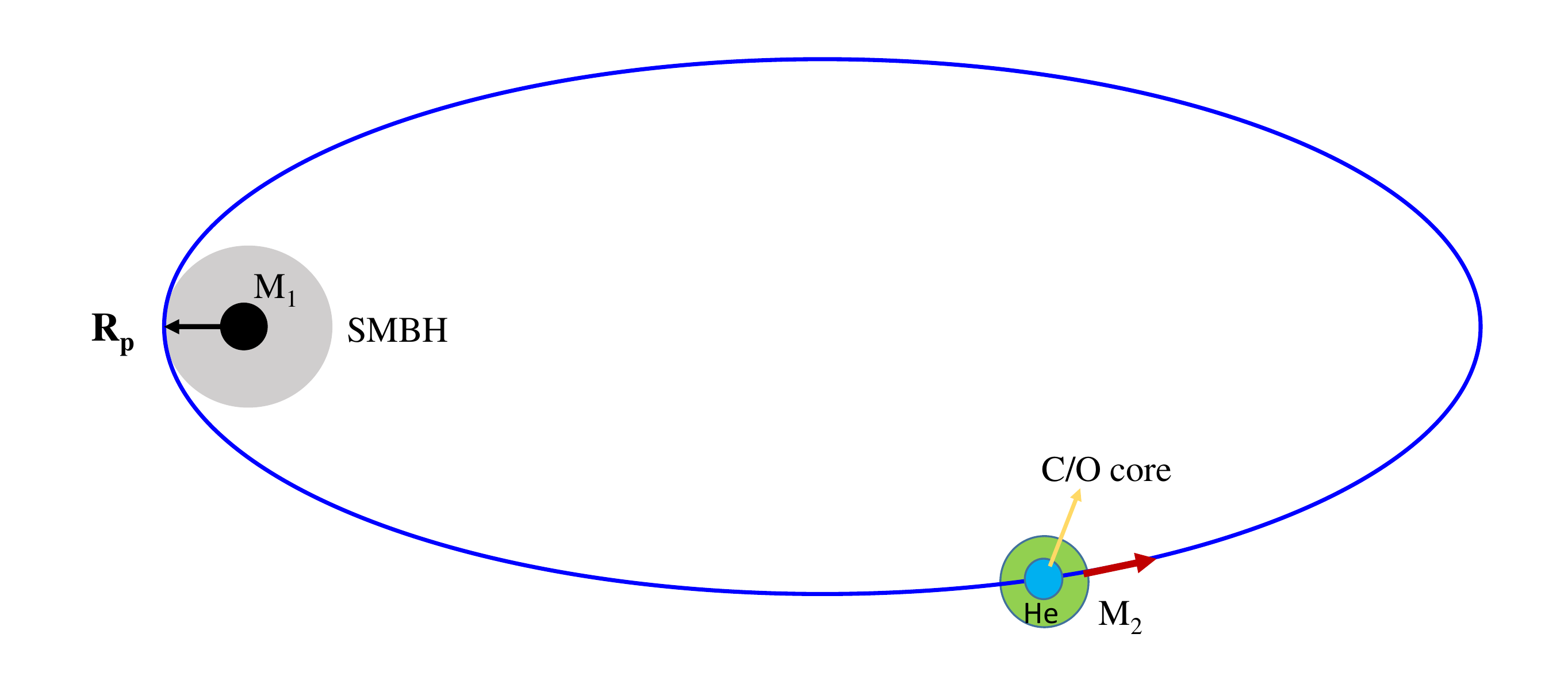}
	\caption{Schematic diagram of the mechanism for QPEs. A hydrogen-deficient post-AGB star with He envelope orbits around an SMBH on an eccentric orbit. When the star approaches periapsis, it just fills its Roche lobe. The He envelope with low density will be partially tidally stripped by the SMBH, while the compact core will survive during the encounter. Finally, the stripped material will fall back to the SMBH, resulting in QPEs. }
	\label{fig:model}
\end{figure}

We consider an SMBH with hydrogen-deficient post-AGB star orbiting around it on an eccentric orbit, as is illustrated in Figure \ref{fig:model}. It has a compact core surrounded by He envelope. When the star approaches periapsis, it marginally fills its Roche lobe, i.e., the star’s radius $R_2$ equals its Roche lobe size. The He envelope with low density will be tidally stripped by the SMBH, while the compact core will survive during the encounter. The stripped matter will lost the orbital energy during the circularization stage. Finally, in the case of highly eccentric orbit, about half of the stripped material will fall back to the SMBH, which produces X-ray eruptions. At each periapsis passage,  mass-transfer onto the SMBH will generate QPEs. Mass transfer from the companion onto the SMBH is driven by angular
momentum loss caused by gravitational radiation.  The system can be analogous to an eccentric version of short-period
cataclysmic variable evolution \citep{Paczynski_1981}. 

The observed recurrence time of QPEs is the orbital period $P$ in our model, and the orbital semimajor axis is
\begin{equation}\label{a}
	a=\left(\frac{G M_1 P^{2}}{4 \pi^{2}}\right)^{1 / 3}=1.6\times10^{12} \ \mathrm{cm}\left( \frac{M_1}{10^{5} \  M_{\odot}}\right)^{1/3}\left( \frac{P}{1 \ \mathrm{h}}\right)^{2/3}.
\end{equation}
In this section, strict constraints of our model are given. In order to produce luminous QPEs with fast rise times, at least three requirements should be satisfied. 
First, the star’s radius equals its Roche lobe size at periapsis. Second, the stars need to be dense enough to make the fast rise timescale of QPEs. Third, the mass-transfer rate driven by gravitational radiation loss should be large enough to generate the luminosities of QPEs.

\subsection{The stable radius for mass-loss}
\begin{figure*}
	\centering
	\includegraphics[width=0.6\textheight]{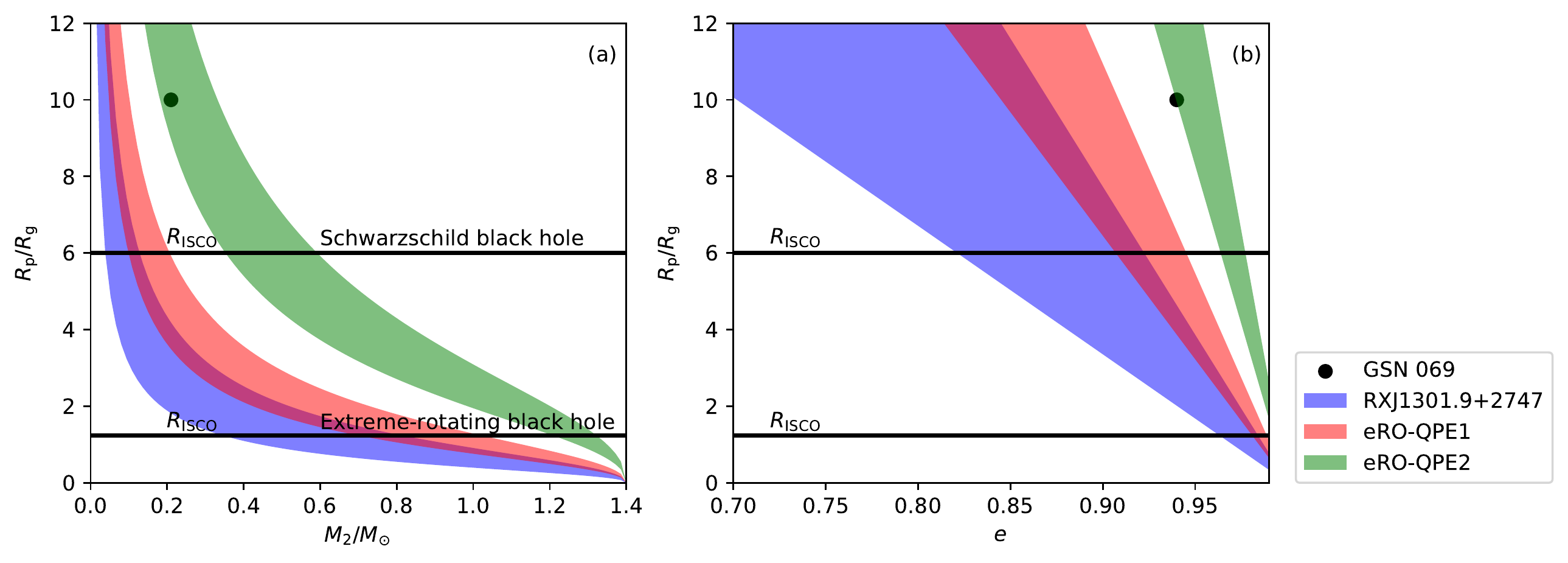}
	\caption{Periapsis $R_{\text{p}}$ in units of the gravitational radius from the mass–radius relation of WDs (shown in panel (a)) and orbital eccentricity (shown in panel (b)). The results of \citet{King_2020} are shown in black circles for GSN 069. For eRO-QPE1 and eRO-QPE2, $M_1$ is estimated by $0.025\% M_{\text {stellar}}$ \citep{Reines_2015}. The radii of ISCO for Schwarzschild BHs ($a=0$) and extreme-rotating BHs ($a=0.998$) are also illustrated in black lines. Panel (a): The mass–radius relation of WDs is taken from \citet{Zalamea_2010}. The case of WDs companion for RXJ1301.9+2747 and eRO-QPE2 is only possible for extreme low-mass WDs or Kerr BHs. Panle (b): Unless the eccentricity is very high, the mass transfer can be carried out stably.}
	\label{Rp}
\end{figure*}

The star should fill the Roche lobe at periapsis $R_{\text{p}}$. The radius of the Roche lobe $R_{\text{lobe}}$ is taken from \cite{Sepinsky_2007}. We can get
\begin{equation}\label{R_lobe}
	R_{2}=R_{\text {lobe }} \simeq 0.46\left(\frac{M_{2}}{M_{1}}  \right)^{1 / 3} R_{\text{p}}.
\end{equation}
In this case, $\beta \simeq0.46$. We discuss the orbital stability of the onset of mass-loss. For a Schwarzschild BH, the radius of innermost stable circular orbit (ISCO) is 
\begin{equation}
	R_{\mathrm{ISCO}}=\frac{6 G M_{\mathrm{1}}}{c^{2}} \simeq 8.86 \times 10^{10} M_{1,5} \mathrm{~cm},
\end{equation}
where $M_{1,5}=M_{1}/10^{5} M_{\odot}$. For Kerr BHs, the corresponding radius of ISCO are
\begin{equation}
	\begin{aligned}
		&R_{\mathrm{ISCO}}=\frac{G M_{\mathrm{1}}}{c^{2}}\left\{3+Z_{2} \mp\left[\left(3-Z_{1}\right)\left(3+Z_{1}+2 Z_{2}\right)\right]^{1 / 2}\right\} \\
		&Z_{1} \equiv\left(1-3 a_{\text{s}}^{2} / M_{\mathrm{1}}^{2}\right)^{1 / 3}\left[\left(1+a_{\text{s}} / M_{1}\right)^{1 / 3}+\left(1-a_{\text{s}} / M_{1}\right)^{1 / 3}\right] \\
		&Z_{2} \equiv\left(3 a_{\text{s}}^{2} / M_1^{2}+Z_{1}^{2}\right)^{1 / 2}
	\end{aligned}
\end{equation}
for co-rotating and counter-rotating case respectively, where $a_{\text{s}}$ is the dimensionless spin \citep{Bardeen_1972,Jefremov_2015}. For $a_{\text{s}}=1$ and $a_{\text{s}}=-1$, the $R_{\mathrm{ISCO}}$ is $1R_{\mathrm{g}}$ and $9R_{\mathrm{g}}$ respectively, where $R_{\mathrm{g}}=GM_1/c^2$ is the gravitational radius. The realistic upper limit for $a_{\text{s}}$ is 0.998 \citep{Thorne_1974}, and the radius of ISCO is $R_{\mathrm{ISCO}}\simeq1.237 R_{\mathrm{g}}$. The mass–radius relation of WDs can be well approximated by \citep{Zalamea_2010}
\begin{equation}\label{M-R}
	R_2=R_{\star}\left(\frac{M_{\mathrm{Ch}}}{M_2}\right)^{1 / 3}\left(1-\frac{M_2}{M_{\mathrm{Ch}}}\right)^{0.447}
\end{equation}
for $0.2 M_{\odot}<M_2<1.4M_{\odot}$, where $M_{\mathrm{Ch}}$ is the Chandrasekhar mass and $R_{\star}=0.013R_{\odot}$. From Equations (\ref{R_lobe}) and (\ref{M-R}), we can get periapsis radius where is the onset of mass-loss for four QPEs (see Figure \ref{Rp}). The mass of SMHB of GSN 069 is $M_1 \sim 4 \times 10^5 \ M_{\odot}$ \citep{Miniutti_2019}, and for RXJ1301.9+2747 is $M_1 \sim (0.8-2.8) \times 10^6 \ M_{\odot}$ \citep{Giustini_2020}. Although, the masses of the SMBHs for eRO-QPE1 and eRO-QPE2 are unknown, we can use the BH-to-total stellar mass fraction $M_{\mathrm{1}} / M_{\text {stellar }} \sim 0.025 \%$ \citep{Reines_2015} to estimate the mass of the central SMBH. The total stellar masses of the host galaxies for eRO-QPE1 and eRO-QPE2 are found to be $3.8_{-1.9}^{+0.4} \times 10^{9} M_{\odot}$ and $1.01_{-0.50}^{+0.01} \times 10^{9} M_{\odot}$ \citep{Arcodia_2021}, respectively. Thus, the mass of the SMBH of eRO-QPE1 and eRO-QPE2 is $\sim 10^{5}-10^{6} \ M_{\odot}$. The result of \cite{King_2020} associated with a low-mass WD companion are shown in black circle. For high-mass SMHBs, the mass of the WD companion should be extreme low (see Section \ref{mass-transfer}).  Therefore, the companion cannot be a WD.

\subsection{The mean density of the companion}
The light curves of the QPEs have a fast rise and a slow decay \citep{Miniutti_2019,Arcodia_2021}. The decay timescale is determined by the viscous timescale of accretion disk, and the rise timescale relates to the radiation region or the interaction between the stripped material and the disk \citep{Shen_2019}. The rise time is comparable to the local circularly orbital timescale at periapsis
\begin{equation}
		t_{\mathrm{cir}}\left(R_{\mathrm{p}}\right)=2 \pi \sqrt{\frac{R_{\mathrm{p}}^{3}}{G M}} \simeq 18 t_{\mathrm{dyn}},
\end{equation}
where $t_{\mathrm{dyn}} \simeq (G\bar{\rho}_{2})^{-1/2}$ is the dynamical timescale of the companion star \citep{Shen_2019}. The local circularly orbital timescale at periapsis is used to estimate the rise timescale.

For CO WDs ($t_{\mathrm{dyn}} \simeq 3$ s), the rise time of the flux is estimated as $t_{\mathrm{rise}}\sim 1$ min. For He WDs  with the mass of 0.2 $M_{\mathrm{\odot}}$($t_{\mathrm{dyn}} \simeq 12$ s), the rise timescale is $t_{\mathrm{rise}}\sim 0.06$ h. \cite{Law-Smith_2017} had constructed a He WD with hydrogen envelope via MESA stellar evolution code, and they found the dynamical timescale is $t_{\mathrm{dyn}} \simeq 535$ s. From the phase-folded light curve of QPEs, the rise timescale is $\sim 0.2-3$ h. The rise timescale of QPEs is much longer than that of WDs. For short duration QPEs (e.g., GSN 069, RXJ1301.9+2747 and eRO-QPE2), the rise timescale is much shorter than He WDs with hydrogen envelopes. From equation (6), we can derive 52 g cm$^{-3}$ $\lesssim\bar{\rho}_{2}\lesssim 4 \times 10^4$ g cm$^{-3}$ from 12 s $\lesssim t_{\mathrm{dyn}} \lesssim$ 535 s. To sum up, the fast rise timescale requires that the stripped material is much denser than Sun-like stars ($\bar{\rho}\simeq 1\ \mathrm{g \ cm^{-3}}$) or red giants ($\bar{\rho}\simeq 10^{-4} \ \mathrm{g \ cm^{-3}}$).

The MS stars, red giants and WDs cannot meet the mean density discussed above. The evolved or stripped stars which lost their H envelopes will be dense enough. In this work, we focus on the evolved stars. There are two main evolving channels of the hydrogen-deficient stars. For low-or-intermediate-mass stars (initial mass 1 $M_\odot<M<$10 $M_\odot$),  H envelope was removed in VLTP phase \citep{Herwig1999} (see Section \ref{stars}). The other case is the product of the evolution of massive stars (e.g., Wolf-Rayet stars). We do not consider massive stars for the following reasons. First, the massive stars have more obvious optical/UV variability during the circularization stage, but all detected QPEs lack the optical/UV counterpart \citep{Miniutti_2019,Giustini_2020,Arcodia_2021}. Second, the strong emission bands of helium were found for Wolf-Rayet stars \citep{Hiltner_1966,Sander_2012}, which is inconsistent with the optical observations of eRO-QPE1 and eRO-QPE2 \citep{Arcodia_2021}. Third, this kind of massive stars is rare.

\subsection{Orbital eccentricity}\label{e}
The tidal stripping process is stable for an EMRI binary system lasting for $\sim$10 000-100 000 years when the star fills the Roche lobe \citep{Hameury_1994,Dai_2013}. The star should fill the Roche lobe at periapsis.  From Equations (\ref{a}) and (\ref{R_lobe}), we can get the relation between the star parameters and the eccentricity
\begin{equation}
	0.46\left(\frac{G P^{2}}{4 \pi^{2}}\right)^{1 / 3} M_{2}^{1 / 3}(1-e)=R_{2},
\end{equation}
or in terms of the average density
\begin{equation}\label{rho_e}
	0.46\left(\frac{G P^{2}}{3 \pi}\right)^{1 / 3} \bar{\rho}_{2}^{1 / 3}(1-e)=1.
\end{equation}
Obviously, the orbital eccentricity must be positive ($e>0$), so  
\begin{equation}\label{rhomin}
	\bar{\rho}_{2}>1.1 \times 10^{2} \mathrm{g \ cm}^{-3}\left(\frac{P }{1\ \mathrm{h}}\right)^{-2}.
\end{equation}
From the $R_{\text{p}}>R_{\text{ISCO}}$ for a Schwarzschild BH, the eccentricity needs to meet
	\begin{equation}
		1-e>0.055 \left(\frac{M_{1}}{10^{5}\ M}_{\odot}\right)^{\frac{2}{3}}\left(\frac{P}{1\ \mathrm{h} }\right)^{-\frac{2}{3}}.
	\end{equation}
Periapsis radius $R_{\text{p}}$ as a function of eccentricity for four QPEs are shown in Figure \ref{Rp}. For GSN 069, $e=0.94$ is taken from \cite{King_2020}, which is shown as black circle.  For eRO-QPE1 and eRO-QPE2, $M_1$ is estimated by $0.025\% M_{\text {stellar}}$ \citep{Reines_2015}. For high eccentricity orbits, the mass transfer is not stable, i.e., the WD companion will be swallowed by the SMBH ($R_{\text{p}}<R_{\text{ISCO}}$).

\subsection{Mass-transfer rate}\label{mass-transfer}
The accretion luminosity of the stripped material falling back to the SMBH is $L=\epsilon \dot{M} c^{2}$, where $\epsilon$ is the radiative efficiency. We take a typical value $\epsilon=0.1$ which has been widely used, such as \cite{King_2020}. In order to generate luminous QPEs, the cycle-average mass-accretion rate is
\begin{equation}\label{dMdt}
	\dot{M}_{2}=\eta\frac{L_{p} }{\varepsilon c^{2}}\simeq 3.5 \times 10^{-5}\ M_{\odot}\ \mathrm{yr}^{-1} \left(\frac{\eta}{0.2}\right)\left(\frac{\varepsilon}{0.1}\right)^{-1}\left(\frac{L_{p}}{ 10^{42}\  \mathrm{erg \ s^{-1}}}\right),
\end{equation}
where $\eta=\Delta P/P$ is the average duty cycle. 

Similar to CV binary evolution \citep{Paczynski_1981}, the mass transfer is driven by angular momentum loss caused by gravitational radiation. The gravitational wave radiation is efficient due to the short period and it will drive the the mass transfer near periapsis. The orbital angular momentum lost caused by gravitational radiation is \citep{Paczynski_1981}
\begin{equation}
	\frac{\dot{J}}{J}=-\frac{32}{5} \frac{G^{3}}{c^{5}} \frac{M_{1} M_{2} M}{a^{4}} f(e),
\end{equation}
where $M=M_{1}+M_{2}$ is the total mass, and 
\begin{equation}
	f(e)=\frac{1+\frac{73}{24} e^{2}+\frac{37}{96} e^{4}}{\left(1-e^{2}\right)^{7 / 2}}.
\end{equation}
In the case of the extreme eccentricity $e \to 1$, $f(e)\simeq 2^{-7/2}(1+73/24+37/96)(1-e)^{-7/2} $. As discussed by \cite{King_2020},  periapsis separation is almost constant. Therefore, the variation of the orbital angular momentum is mainly resulted from the mass transfer of the secondary star $\dot{J}/J=\dot{M}_{1}/M_{1}+\dot{M}_{2}/M_{2}=\dot{M}_{2}/M_{2}\times (1-M_2/M_1) \simeq \dot{M}_{2}/M_{2}$ \citep{King_2020}. Therefore, we can get the mass transfer rate due to gravitational wave radiation as
\begin{equation}\label{dmdt}
	\dot{M}_{2}\simeq 5.4 \times 10^{-5}M_{\odot}\ \mathrm{yr}^{-1}\left(\frac{M_{1}}{10^{5}\ M_{\odot}}\right)^{\frac{2}{3}}\left(\frac{M_{2}}{1\ M_{\odot}}\right)^{2}\left(\frac{P}{1\ \mathrm{h} }\right)^{-\frac{8}{3}}(1-e)^{-7/ 2}.
\end{equation}

\begin{figure*}
	\centering
	\includegraphics[width =0.8\textwidth]{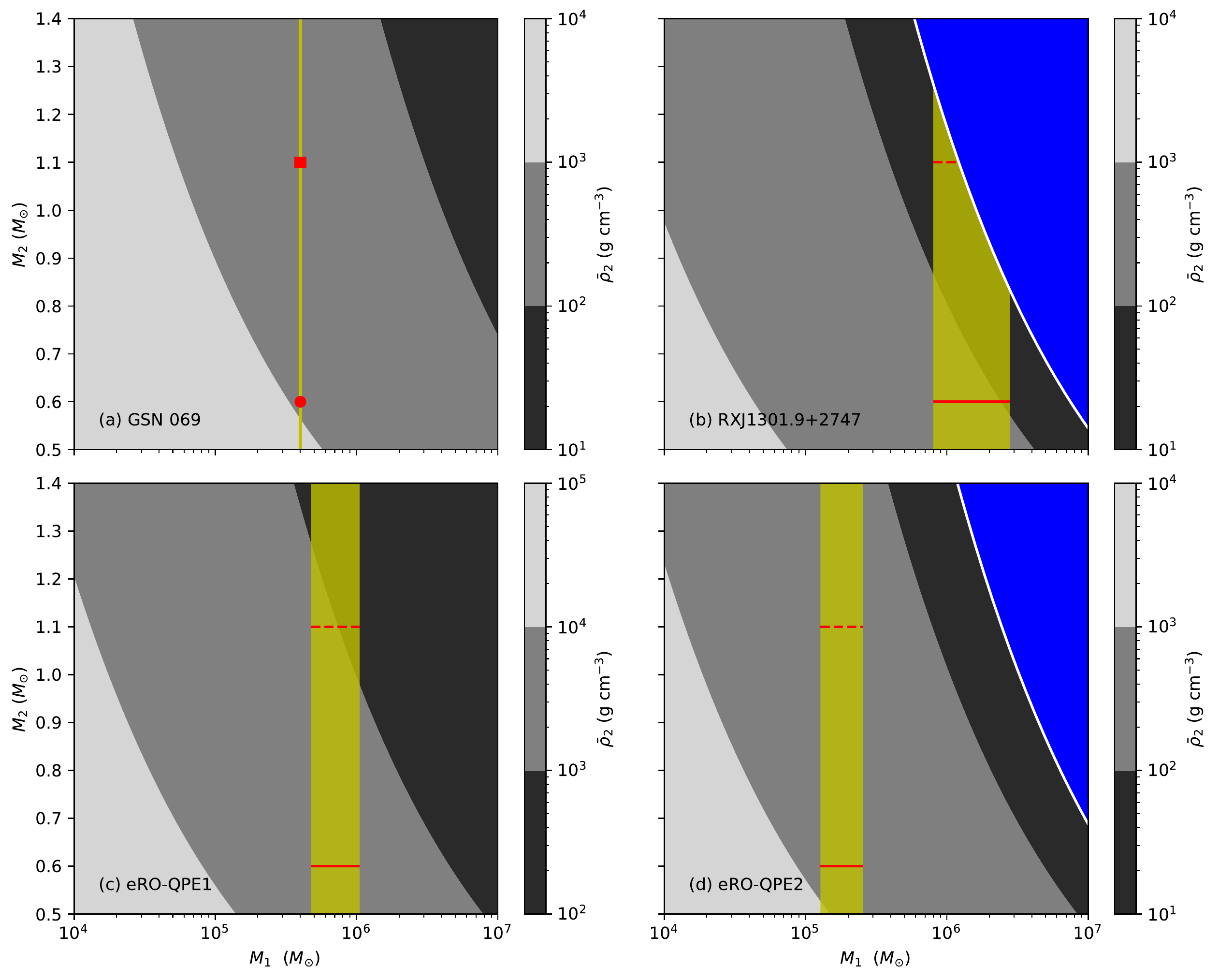}
	\caption{Contours of allowed average density for different values of the SMBH mass $M_1$ and the companion star mass $M_2$. The white solid lines refer to constraints from the rise time. The excluded region is shown in blue. The gray region shows different densities of the companion star. The yellow vertical solid line (or panel (a)) and shading regions (for panel (b), (c) and (d)) represent the mass of SMBHs. The allowed parameter space is shown in yellow. For GSN  069,  the star with mass of 0.6 $M_{\odot}$ and 1.1 $M_{\odot}$ are shown in red circle and square in panel (a), respectively. For the other three shown in panel (b), (c) and (d), the star with mass of 0.6 $M_{\odot}$ and 1.1 $M_{\odot}$ are shown in red solid and dashed lines, respectively. The average density of stars to explain the properties of QPEs is listed in Table \ref{table2}, which is between 52.34 g cm$^{-3}$ and 3642.22 g cm$^{-3}$. Only low and intermediate mass stars which H envelopes are lost in the VLTP phase \citep{Herwig1999} can fulfill these requirements. The mass of hydrogen-deficient post-AGB stars are almost the same as the final stellar evolution product WDs, but the density is lower because the existence of He envelopes.}
	\label{rho-M1-M2}
\end{figure*}

The mass-transfer time-scale is $t_{\text{GW}}=-M_2/\dot{M}_2\sim10^5$ yr. If the post-AGB stars which still have burning He shells are captured by the SMBHs, the onset of RLOF would also be caused by nuclear evolution. The nuclear timescale of He shell burning is $10^5-10^6$ yr, which is similar to $t_{\text{GW}}$.  The ongoing RLOF process caused by thermodynamical conditions of the convective envelope is also discussed in Section \ref{stars}, and we find the mass transfer rate is mainly determined by gravitational wave radiation. 

From equations (\ref{rho_e}) and (\ref{dmdt}), the mass-transfer rate is
\begin{equation}\label{dm2dt}
	\begin{aligned}
		\dot{M}_{2}\simeq &2.1 \times 10^{-5}M_{\odot}\ \mathrm{yr}^{-1}\left(\frac{M_{1}}{10^{5}\ M_{\odot}}\right)^{\frac{2}{3}}\left(\frac{M_{2}}{1\ M_{\odot}}\right)^{2}\\
		&\times 
		\left(\frac{P}{1\ \mathrm{h} }\right)^{-\frac{1}{3}}\left(\frac{\bar{\rho}_{2}}{200 ~\mathrm{g~cm^{-3}}}\right)^{\frac{7}{6}},
	\end{aligned}
\end{equation}
and the luminosity is
\begin{equation}
	\begin{aligned}
			L=&1.2 \times 10^{41}~ \mathrm{erg~s}^{-1}\left(\frac{\varepsilon}{0.1}\right)\left(\frac{M_{1}}{10^{5} M_{\odot}}\right)^{\frac{2}{3}}\left(\frac{M_{2}}{1 M_{\odot}}\right)^{2}\\
				&\times \left(\frac{P}{1~\mathrm{h}}\right)^{-\frac{1}{3}}\left(\frac{\bar{\rho}_{2}}{200 ~\mathrm{g~cm^{-3}}}\right)^{\frac{7}{6}}.
	\end{aligned}	
\end{equation}
For a MS star (e.g., Sun) and a red giant (e.g.,  $M_2=1 ~M_{\odot}$, $R_2=25 ~R_{\odot}$), the characteristic accretion luminosities are $3.6\times10^{38}$ erg s$^{-1}$ and $6.3\times10^{33}$ erg s$^{-1}$, respectively. So they are much lower than those of observed QPEs.

From Equations (\ref{rho_e}), (\ref{dMdt}) and (\ref{dmdt}), we derive the relation between star parameters ($\bar{\rho}_2$ and $M_2$) and BH mass for four QPEs (see Fig. \ref{rho-M1-M2}). The white  solid lines refer to constraints from the rise time. The infeasible range is in blue. The yellow vertical solid line (for GSN  069 shown in panel (a)) and shading regions (for the other three shown in panels (b), (c) and (d)) represent the mass of the SMBH. The allowed parameter space is shown in yellow. It is obvious that MS stars and red giants with higher mass and lower density are located in the infeasible range. Only low and intermediate mass stars whose H envelopes have been removed in the post-AGB phase can fulfill them. The mass of hydrogen-deficient post-AGB stars are almost the same as the final stellar evolution product WDs ($0.5-1.4M_\odot$), but the density is lower because of the existence of He envelopes. In section \ref{stars}, we will show the evolved low and intermediate mass stars ($1M_\odot<M<10 M_\odot$) which have lost H envelopes in the post-AGB phase can explain all the observation properties. Taken the hydrogen-deficient post-AGB stars constructed in Section \ref{stars} as examples, the required averaged density is depicted in Fig. \ref{rho-M1-M2}. For GSN  069,  the stars with masses of 0.6 $M_{\odot}$ and 1.1 $M_{\odot}$ are shown in red circle and square in panel (a), respectively. For the other QPEs shown in panels (b), (c) and (d), the stars with masses of 0.6 $M_{\odot}$ and 1.1 $M_{\odot}$ are shown in red solid and dashed lines, respectively. The average density of stars required to explain the properties of QPEs is listed in Table \ref{table2}, which is between 52 g cm$^{-3}$ and 3642 g cm$^{-3}$.

\section{Hydrogen-deficient post-AGB stars}\label{stars}
For a low-or-intermediate-mass star  (initial mass $1M_\odot<M<10 M_\odot$), the mass loss will become important in the AGB phase \citep{Bloecker1995}. After hydrogen envelopes are lost in the VLTP phase \citep{Herwig1999}, a hydrogen-deficient star forms. The hydrogen-deficient post-AGB stars have a compact core with He envelopes of a few $10^{-2}M_\odot$. The mass of hydrogen-deficient post-AGB stars is almost the same as the final stellar evolution product WDs ($0.5-1.4M_\odot$), but the density is lower because of the existence of He envelopes. When it evolves to the phase that average density satisfying the range given above, it is captured by the SMBH. It fills the Roche-lobe and donates to the SMBH, generating QPEs.

For the stars with the initial mass $1 M_\odot<M<8 M_\odot$, the degenerate CO cores form after core He burning, and the remnants are CO WDs after the residual burning. If the initial stellar mass is higher ($8 M_\odot<M<10 M_\odot$), the core carbon will be ignited. In this work, we take the evolution of the CO WD with $\sim0.6M_{\odot}$ and ONe WD $\sim 1.1M_{\odot}$  as examples. The initial mass is 3.1$M_{\odot}$ and 10 $M_{\odot}$, respectively. The MESA stellar evolution code \citep{Paxton_2011,Paxton_2013,Paxton_2015,Paxton_2018} is used to construct the hydrogen-deficient post-AGB stars. 

\begin{figure}
	\centering
	\includegraphics[width = 0.5\textwidth]{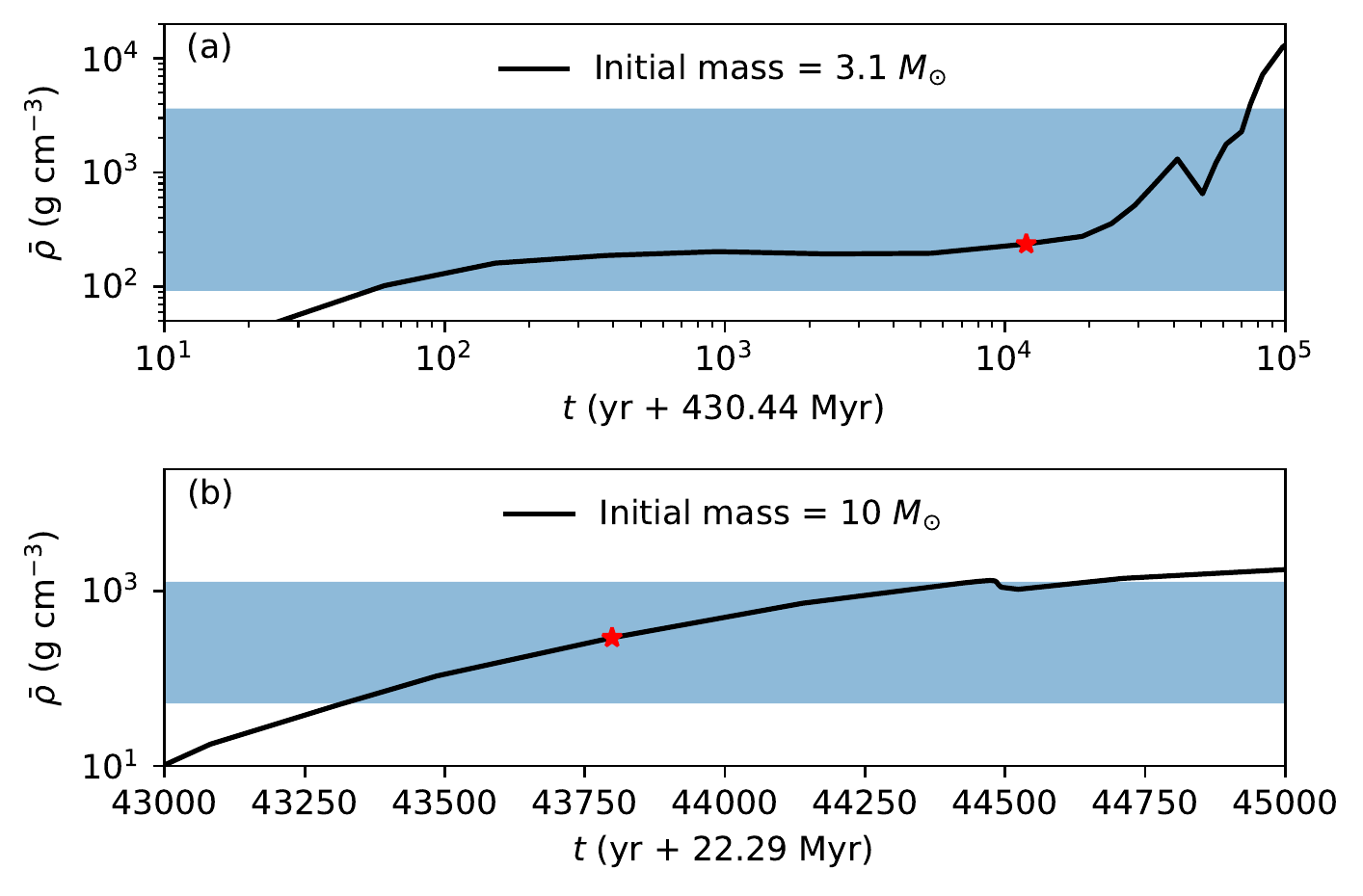}
	\caption{The average density evolution of the post-AGB stars for the initial mass of 3.1 $M_{\odot}$ (shown in panel (a))  and 10 $M_{\odot}$ (shown in panel (b)). The shading regions represent the conditions to produce QPEs given in Table \ref{table2}, which can last for few $10^3$ years and few $10^4$ years for $M_2=0.6$ and $M_2=1.1$, respectively.  The age and average density corresponding to Fig. \ref{M-R} are represented by red stars, and the stellar properties are listed in Table \ref{core}.}
	\label{rho-t}
\end{figure}

\begin{table}
	\centering
	\caption{The required average density of post-AGB stars to produce QPEs.}
	\begin{tabular}{cccc}
		\hline
		\hline
		Source &$M_1$ & $M_2$  &$\bar{\rho}_2$ \\
		            &  ($10^5~M_{\odot}$) &  ($M_{\odot}$) &  (g cm$^{-3}$)   \\
		\hline
		GSN 069 			 & 4 & 0.6 & 899.26\\
		&    & 1.1 & 318.13\\
		RXJ1301.9+2747 & 8$-$28   & 0.6  & 91.81$-$187.85\\
		&            & 1.1  & 52$-$66.46\\
		eRO-QPE1           & 4.75$-$10.5  & 0.6  & 2314.79$-$3642.22\\
		&               & 1.1  & 818.92$-$1288.53\\
		eRO-QPE2           & 1.28$-$2.55              & 0.6  & 538.98$-$800.92\\
		&               & 1.1  & 190.68$-$283.35\\
		\hline
		\hline
	\end{tabular}
	\label{table2}
\end{table}

\begin{figure*}
	\centering
	\includegraphics[width = 0.6\textwidth]{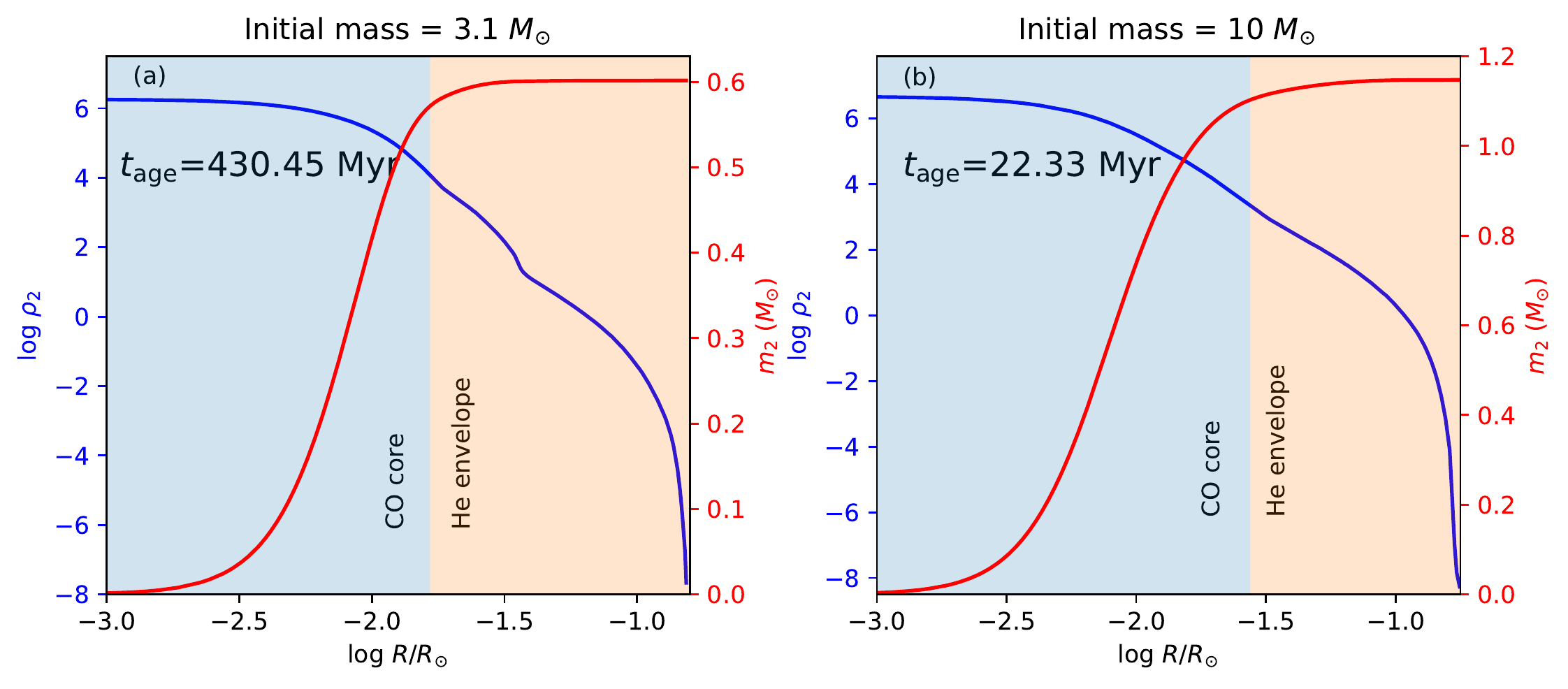}
	\caption{The density (a) and mass (red lines) profiles of the hydrogen-deficient post-AGB stars for the initial mass of 3.1 $M_{\odot}$ (panel a) and 10 $M_{\odot}$ (panel b). The stellar properties are listed in Table \ref{core}. The regions of degenerate CO core and He envelope are shown in blue and orange, respectively.}
	\label{rho-r}
\end{figure*}

In Fig. \ref{rho-t},  the evolution of the average density of hydrogen-deficient post-AGB stars is shown. For the star with an initial mass of 3.1$M_{\odot}$, the degenerate CO core (preformed CO WD) is formed in the final phase of AGB evolution. The black line in the top panel represents the evolution of the post-AGB phase after the H shell extinct in VLTPs. The star has a burning He envelope and finally cools down as a CO WD with the mass of $\sim 0.6M_{\odot}$. However, for the star with an initial mass of 10 $M_{\odot}$, it will become a super-AGB star which is massive enough to ignite C after the giant branch. The black line in the bottom panel represents the evolution in C burning phase after the H envelope was removed due to the strong stellar wind. Finally, the star will become a ONe WD with the mass of $\sim 1.1 M_{\odot}$. The shading regions represent the conditions to produce QPEs given in Table \ref{table2}, which can last for a few $10^3$ years and a few $10^4$ years for $M_2=0.6$ and $M_2=1.1$, respectively.  When the stars evolve onto the post-AGB phase, the stars are captured by the SMBHs. Due to nuclear evolution or gravitational wave radiation, the star marginally fills its Roche lobe at periapsis and donates material to SMBHs, producing QPEs. The density profile of the hydrogen-deficient post-AGB stars is depicted in Fig. \ref{rho-r}. The corresponding age and average density are represented by red stars in Fig. \ref{rho-t}, and the stellar properties are listed in Table \ref{core}. The regions of degenerate CO core and He envelope are shown in blue and orange, respectively.

Because the hydrogen-deficient post-AGB stars we got still have a convective envelope, the mass-transfer caused by thermodynamical conditions is also important \citep{Ritter1988,Kolb1990}. It can be expressed as
\begin{equation}
	\begin{aligned}
		\dot{M}_{0} &=\frac{2 \pi}{\sqrt{e}}\left(\frac{\mathscr{R}  T_{2}}{\mu}\right)^{3 / 2} \frac{R_{2}^{3}}{G M_{2}} \rho_{\text{ph}} F(q)\\
		&=7.66 \times 10^{-9}\ M_{\odot}\ \mathrm{yr}^{-1}\left(\frac{T_{2}}{10^{5} \text{~K}}\right)^{3 / 2}\left(\frac{{\mu}}{4}\right)^{-3 / 2}\\
		&\times \left(\frac{R_{2}}{0.1 ~R_{\odot}}\right)^{3}\left(\frac{M_{2} }{0.6~ M_{\odot}}\right)^{-1}\left(\frac{\rho_{\text{ph}}}{10^{-5} ~\mathrm{g~cm^{-3}}}\right) F(q),
	\end{aligned}
\end{equation}
where $\mathscr{R}$ is the gas constant, $T_2$ is the effective temperature, $\mu$ is the mean molecular weight, $\rho_{\text{ph}}$ is the photosphere density, and $F(q)<<1$ for our interest (e.g., the mass ratio $q=M_1/M_2\sim10^5$).  The characteristic stellar parameters of hydrogen depleted post-AGB stars are given in Table \ref{core}. This mass-transfer rate is negligible compared with that caused by gravitational radiation.

%For the stars with the initial mass 1 M$_\odot<M<$8 M$_\odot$, the degenerate CO cores form after He burning, and the remnants are CO WDs after the envelopes removed. If the initial stellar mass is higher (8 M$_\odot<M<$10 M$_\odot$), the core is non-degenerate and carbon will be ignited. When the H envelopes are removed, a compact CO or ONe core with He envelopes is left. The mass of hydrogen-deficient stars' He envelopes is $M\sim 0.02-0.05$ M$_\odot$. The average density is $\sim 10^2-10^3$ g cm$^{-3}$ (see red dashed lines in Figure \ref{wd}), which is consistent with the constrains from Figure \ref{rho-M1-M2}. Finally, the stars will evolve to CO or ONe WDs with high density (see green dashed–dotted lines in Figure \ref{wd}).

The star is possible to be captured before the post-AGB phase, such as the formation channel of \cite{King_2020}. The SMBH captures a rad giant and strips the H envelope, the left He core (low-mass WD) which donates material to the SMBH and generates QPEs, which was proposed to explain GSN 069. However, the WD companion will fast lose all material after a few thousand orbits \citep{Zalamea_2010}, which can not explain the long-standing QPE \citep{Giustini_2020}. From the  mass-radius relation of WDs \citep{King_2020}, we find that the WDs mass is $\sim 0.05 M_{\odot}$  for RXJ1301.9+2747 and eRO-QPE2. However, the extreme low-mass WDs are rare in the Montreal White Dwarf Database\footnote{ \url{https://www.montrealwhitedwarfdatabase.org/references.html}}\citep{Dufour2017,Kuerban2020}.  In addition, eRO-QPE1 and eRO-QPE2 were discovered in quiescent galactic nuclei, which implies the companion will not fill the Roche-lobe before the post-AGB phase. 

%Using the Sloan Digital Sky Survey Data Release 4, the mean mass of white dwarfs is 0.613 M$_\odot$, and the lowest mass is about 0.2 M$_\odot$ \citep{Tremblay2011}. There is a growing observations for extremely low-mass WDs \citep{Kaplan2014,Brown2016}. The lower mass limit is about 0.15 M$_\odot$ \citep{Kaplan2014,Brown2016}. 

If the star fills its Roche lobe at periapsis, the stripped matter will fall back to the black hole and generate QPEs. From Equation (\ref{dm2dt}), the life span of QPEs can be estimated by
\begin{equation}\label{tau}
	\begin{aligned}
		\tau_{\mathrm{QPE}}=\frac{M_{\text{He}}}{\dot{M}_{2}}&=1428.5 \text { yr } \left(\frac{M_{\text{He}}}{0.03 M_{\odot}}\right)\left(\frac{M_{1}}{10^{5}\ M_{\odot}}\right)^{-\frac{2}{3}}\left(\frac{M_{2}}{1\ M_{\odot}}\right)^{-2}\\
			&\times \left(\frac{P}{1\ \mathrm{h} }\right)^{\frac{1}{3}}\left(\frac{\bar{\rho}_{2}}{200 ~\mathrm{g~cm^{-3}}}\right)^{-\frac{7}{6}}.
	\end{aligned}
\end{equation}
For $M_1=10^{5}\ M_{\odot}$ and $P=1$ h, the life span of QPEs for the post-ABG star donor is $\sim 2721.57$ yr and $\sim 768.40$ yr for $M_2=0.6M_{\odot}$ and $M_2=1.1M_{\odot}$, respectively. The star properties are the same as those in Table \ref{core}. The subsequent evolution of the hydrogen-deficient post-AGB stars can be ignored because the life span of QPEs are quite short. Finally, the He enevlopes are all stripped, the compact objects will be swallowed by the SMBHs.

\begin{table*}
	%\resizebox{\textwidth}{!}{
	\centering
	\caption{The properties of hydrogen-deficient post-AGB stars with initial masses $3.1~M_\odot$ and $10$ $M_\odot$.}
	\begin{tabular}{ccccccccc}
		\hline
		\hline
		$M_{\text{initial}}$  & Age  & $M_{\text{core}}$  & $R_{\text{core}}$ & $M$  & $R$  & log $T_{\text{eff}}$&log $\rho_{\text{ph}}$ &log $\bar{\rho}_2$  \\
		($M_{\odot}$) &(Myr) & ($M_{\odot}$) & ($R_{\odot}$) & ($M_{\odot}$) & ($R_{\odot}$) & (K) & (g cm$^{-3}$) &(g cm$^{-3}$)\\
		\hline
		3.1 & 430.31 & 0.57 & 0.017 & 0.60 & 0.15 & 4.97 & -7.67 & 2.37 \\
		10 & 22.33 & 1.10 & 0.027 & 1.14 & 0.18 & 5.18 & -8.27 & 2.47 \\
		\hline
		\hline
	\end{tabular}
	%}
	
	\label{core}
\end{table*}

\section{Event rate}\label{event_rate}
The standard formation channel of EMRIs is the capture of a compact object (WD, NS or BH) by an SMBH \citep{Sigurdsson_1997,Amaro-Seoane_2007,Amaro-Seoane_2018}. Its rate is about a few percent of the TDE rate. Some other processes include tidal separation of compact binaries,
formation or capture of massive stars in accretion disks \citep{Amaro-Seoane_2007,Maggiore:2018sht}. In addition, ``fake plunges'' can serve as high-eccentric EMRIs with a rate about 30 times larger than the typical rate of EMRIs \citep{Amaro-Seoane_2013}. Considering the stars injected on high-eccentric orbits in the vicinity of the SMBH due
to the Hills binary disruption, the EMRI rate can approach the TDE rate if the binary fraction at the SMBH affecting radius is close to unity \citep{Sari_2019}. Interestingly, the fraction of binaries is larger than 50\% from observations. Therefore, we safely assume that the total EMRI rate has the same order as the TDE rate.
Below, we follow the method proposed by our previous work \citep{Wang_2019} to estimate the QPE rate. The mass of SMBHs can be approximated by $M_{\text{BH}}$-$\sigma$ relation
\begin{equation}
	M_{\mathrm{BH}}=M_{\mathrm{BH}, *}\left(\frac{\sigma}{\sigma_{*}}\right)^{\lambda}
\end{equation}
where $\sigma$ is the spheroid velocity dispersion. The $M_{\text{BH}}-\sigma$ relation also applies for low-mass SMBH ($<10^6 M_{\odot}$), considering the uncertainties \citep{Xiao_2011}. Hence, this relation is used in this work. Combined with the constraints from galaxy luminosity functions and the $L$–$\sigma$ correlation \citep{Aller_2002}, the BH mass function is \citep{Gair_2004}
\begin{equation}
	M_{\mathrm{BH}} \frac{\text{d} N}{\text{d} M_{\mathrm{BH}}}=\phi_{*} \frac{\epsilon}{\Gamma\left(\frac{\gamma}{\epsilon}\right)}\left(\frac{M_{\mathrm{BH}}}{M_{\mathrm{BH}, *}}\right)^{\gamma} \times \exp \left[-\left(\frac{M_{\mathrm{BH}}}{M_{\mathrm{BH}, *}}\right)^{\epsilon}\right]
\end{equation}
where $\sigma=3.08/\lambda$, $\phi_*$ is the total number density of galaxies,
and $\Gamma(z)$ is the gamma function. The spatial density of BHs can be estimated from the parameters of low-mass SMBHs ($<10^6 M_{\odot}$) \citep{Aller_2002}
\begin{equation}
	M_{\mathrm{BH}} \frac{\text{d} N}{\text{d} M_{\mathrm{BH}}}=2 \times 10^{-3} h_{70}^{2} \mathrm{Mpc}^{-3}
\end{equation}
where $h_{70} \equiv H_0/70$ km s$^{-1}$ Mpc$^{-1}$ is the dimensionless Hubble constant. Then, for solar-type stars, the disruption rate per galaxy is \citep{Wang_2004}
\begin{equation}
	\label{rate}
	\mathcal{R}=6.5 \times 10^{-4} \mathrm{yr}^{-1}\left(\frac{M_{*}}{M_{\odot}}\right)^{-1 / 3} \times\left(\frac{R_{*}}{R_{\odot}}\right)^{1 / 4}\left(\frac{M_{\mathrm{BH}}}{10^{6} M_{\odot}}\right)^{-1 / 4}
\end{equation}
The number ratio of 1-10 $M_{\odot}$ stars to solar-type stars is about 0.48 using the Salpeter initial mass function, while the lifetime ratio is estimated to be $10^{-2}$ by average. Since the density of the He envelope in our scenario is $10^2$ to $10^3$ times larger than that of solar-type stars, resulting in a smaller tidal radius. We reduce the rate by a factor of $1\times 10^{-2}$. Last but not least, our model requires the star lying in the He main sequence, whose duration is roughly 0.1 times that of the H main
sequence. Integrating equation (\ref{rate}) over $1 M_{\odot} < M_* < 10 M_{\odot}$, $20 R_{\odot} < R_* < 60 R_{\odot} $ and combining all the fore-mentioned factors gives the event rate $\dot{N}_{\text{QPE}}=$1.5 Gpc$^{-3}$ yr$^{-1}$ for $M_{\text{BH}} = 5\times 10^5 M_{\odot}$. Therefore, the observed number of QPEs can be calculated as
\begin{equation}
	N_{\text{QPE}} \sim \dot{N}_{\text{QPE}}V\tau_{\text{QPE}},
\end{equation}
where $V$ is the searching volume, and $\tau_{\text{QPE}}$ is the active lifttime of QPEs.
The co-moving volume within the redshift $z=0.0505$ of
the most distant QPE event (eRO-QPE1) is $V\sim 0.04$Gpc$^3$.
In our model, the range of $\tau_{\text{dest}}$ is $100 \sim 1000 \text{yr}$ (equation (\ref{tau})). Hence, $N_{\text{QPE}}$ is between $6- 60$. It has been
estimated that \textit{eROSITA} would discover up to about 10 or 15 QPEs by the end of 2023 \citep{Arcodia_2021}, which is well consistent with our estimation.

\section{Gravitational wave signals detection}\label{gw}
The mass-loss systems involving MS and SMBH can be the GW sources \citep{Linial2017}. In our model, these QPEs are also promising GW sources for space-based GW detectors, such as LISA \citep{Amaro-Seoane_2017,Amaro-Seoane_2018} and TianQin Project \citep{Luo_2016}. The Keplerian orbital frequency of QPEs is about $ f_{\text{orb}}\sim10^{-4}$ Hz. The compact core with helium envelopes inspiral into the SMBH will produce EMRI signals at frequency $f=2f_{\text{orb}}$, which can be detected by LISA and Tianqin. The current LISA mission, planned to be launched in 2030s,  has an arm-length of $2.5 \times 10^9$ m and is sensitive to low frequency bands ($10^{-4} \sim 10^{-1}$ Hz). The Tianqin has a similar scientific goal, but uses earth orbit instead of heliocentric orbit. Generally, it takes several years for the compact core to plunge into the SMBH after entering the GW detection bands. Therefore, $10^4$ to $10^5$ circles can be recorded by detectors to build up the signal-to-noise ratio (SNR) with hierarchical matched filtering method, which divides data into short segments and adds their power incoherently. 

\begin{figure}
	\centering
	\includegraphics[width = 0.5\textwidth]{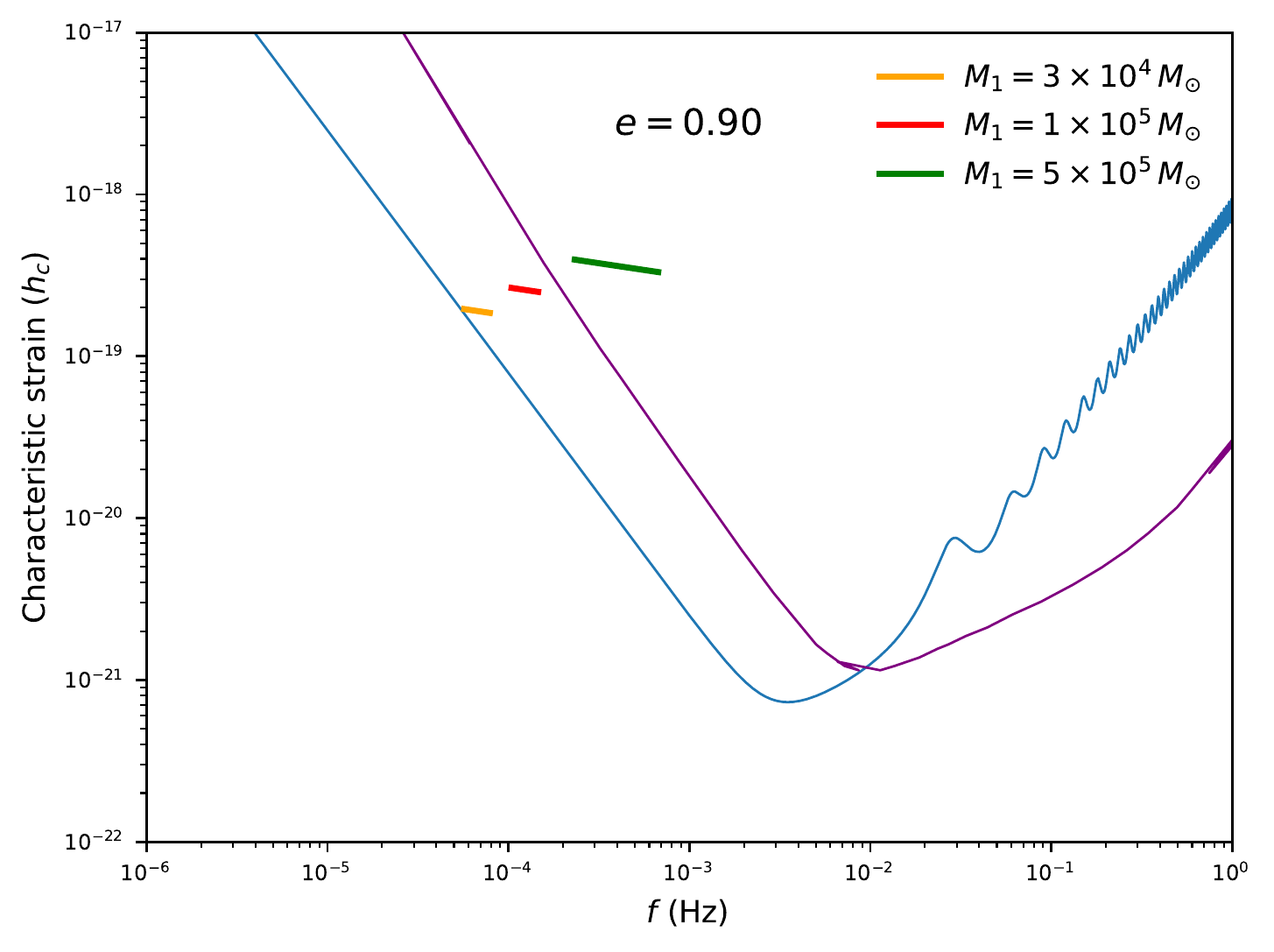}
	\caption{The diagram of the characteristic strains of EMRIs for different BH masses. We set $z=0.02$, $M_2=0.6 M_{\odot}$ , $a=5\times 10^{12}$ cm and $e=0.9$. The sensitivity curves of LISA (blue line) and Tianqin (purple line) are also plotted for comparison. These QPEs are promising EMRI sources for LISA and Tianqin.}
	\label{GW}
\end{figure}

The GW emission power evolution is \citep{Peters_1964}
\begin{equation}
	\dot{E}=-\frac{32}{5} \frac{G^{4} M_{1}^{2} M_{2}^{2} M}{c^{5} a^{5}}f(e),
\end{equation}
and the Keplerian orbital evolution is
\begin{equation}
	\dot{a}=-\frac{64}{5} \frac{G^{3} M_{1} M_{2} M}{c^{5} a^{3}}f(e).
\end{equation}
The characteristic strain of the GW emission from the proper distance $D$ away from the detector is \citep{Amaro-Seoane_2018,Maggiore:2018sht} 
\begin{equation}
	h_{c}(f)=2 f|\tilde{h}(f)|=\left(\frac{2 f^{2}}{\dot{f}}\right)^{1 / 2} h_{0}=\frac{(2 \dot{E} / \dot{f})^{1 / 2}}{\pi D},
\end{equation}
where $h_0$ is an instantaneous root-mean-square amplitude. Unlike \citet{Chen2021}, GW radiation in harmonic frequencies is not considered. The characteristic strains of different BH masses are shown in Figure \ref{GW}, where we set $z=0.02$, $M_2=0.6 M_{\odot}$, $a=5\times 10^{12}$ cm and $e=0.9$. These GW signal of these QPEs is well above the sensitivity curves of LISA (blue line) and Tianqin (purple line). 
These sources with EMRI signals and electromagnetic counterparts are important for cosmological purposes, such as measuring the Hubble constant \citep{Abbott_2017,Chen_2018,Yu2018} and the peculiar velocity \citep{Wang_2018,Palmese2021}.

\section{Summary}\label{conclusion}
In this paper, we have proposed a hydrogen-deficient post-AGB star orbiting the SMBH as the generation mechanism of QPEs. The whole picture is as following. A star with an initial mass between 1-10 $M_\odot$ evolves into a post-AGB phase, then it is captured by a SMBH to form an elliptic orbit. When it passes the periapsis, the star fills its Roche lobe, leading to mass transfer to the SMBH. The accretion of the mass by the SMBH will produce QPEs. 

According to the rise time,  the orbit stability and the luminosity of QPEs, we find the average density of the companion ranges from tens to thousands of g cm$^{-3}$. The average density of the donor of GSN 069, RXJ1301.9+2747, eRO-QPE1and eRO-QPE2 is expected to be 899.26 g cm$^{-3}$, 91.81$-$187.85 g cm$^{-3}$, 2314.79$-$3642.22 g cm$^{-3}$ and 538.98$-$800.92 g cm$^{-3}$ for $ M_{2}=0.6M_{\odot} $, respectively. For $  M_{2}=1.1M_{\odot} $, the required average density is 318.13 g cm$^{-3}$, 52$-$66.46 g cm$^{-3}$, 818.92$-$1288.53 g cm$^{-3}$ and 190.68$-$283.35 g cm$^{-3}$, respectively. The properties of hydrogen-deficient post-AGB stars are consistent with these constraints.

The MESA stellar evolution code is used to construct the evolutions of low and intermediate-mass stars. We find that when they lost the 
H envelopes in the VLTP phase and evolve to hydrogen-deficient post-AGB stars can meet the constraints on the average density. The rate of QPEs is estimated to be 1.5 Gpc$^{-3}$ yr$^{-1}$. In our model, QPEs are promising candidates for the electromagnetic counterparts of EMRIs.  The GWs from these QPEs can be detected by LISA and Tianqin. 

\section*{Acknowledgements}
We thank the anonymous referee for helpful comments, Rong-Feng Shen, Xin-Wen Shu, Ting-Gui Wang, Abudushataer Kuerban and Ying Qin for discussions. This work was supported by the National Natural Science Foundation of China (grant Nos. U1831207 and 11833003), 
the Fundamental Research Funds for the Central Universities
(No. 0201-14380045), and the National Key Research and Development Program
of China (grant No. 2017YFA0402600 and 2020YFC2201400), the National
SKA Program of China (No. 2020SKA0120300) and NWO, the Dutch Research Council, under Vici research programme `ARGO' with project number 639.043.815. 

\bibliographystyle{aa}
\bibliography{ms}

\begin{thebibliography}{64}
\expandafter\ifx\csname natexlab\endcsname\relax\def\natexlab#1{#1}\fi

\bibitem[{{Abbott} {et~al.}(2017){Abbott}, {Abbott}, {Abbott}, {Acernese},
  {Ackley}, {Adams}, {Adams}, {Addesso}, {Adhikari}, {Adya}, \&
  et~al.}]{Abbott_2017}
{Abbott}, B.~P., {Abbott}, R., {Abbott}, T.~D., {et~al.} 2017, \nat, 551, 85

\bibitem[{{Aller} \& {Richstone}(2002)}]{Aller_2002}
{Aller}, M.~C. \& {Richstone}, D. 2002, \aj, 124, 3035

\bibitem[{{Amaro-Seoane}(2018)}]{Amaro-Seoane_2018}
{Amaro-Seoane}, P. 2018, Living Reviews in Relativity, 21, 4

\bibitem[{{Amaro-Seoane} {et~al.}(2017){Amaro-Seoane}, {Audley}, {Babak},
  {Baker}, {Barausse}, {Bender}, {Berti}, {Binetruy}, {Born}, {Bortoluzzi},
  {Camp}, {Caprini}, {Cardoso}, {Colpi}, {Conklin}, {Cornish}, {Cutler},
  {Danzmann}, {Dolesi}, {Ferraioli}, {Ferroni}, {Fitzsimons}, {Gair}, {Gesa
  Bote}, {Giardini}, {Gibert}, {Grimani}, {Halloin}, {Heinzel}, {Hertog},
  {Hewitson}, {Holley-Bockelmann}, {Hollington}, {Hueller}, {Inchauspe},
  {Jetzer}, {Karnesis}, {Killow}, {Klein}, {Klipstein}, {Korsakova}, {Larson},
  {Livas}, {Lloro}, {Man}, {Mance}, {Martino}, {Mateos}, {McKenzie},
  {McWilliams}, {Miller}, {Mueller}, {Nardini}, {Nelemans}, {Nofrarias},
  {Petiteau}, {Pivato}, {Plagnol}, {Porter}, {Reiche}, {Robertson},
  {Robertson}, {Rossi}, {Russano}, {Schutz}, {Sesana}, {Shoemaker}, {Slutsky},
  {Sopuerta}, {Sumner}, {Tamanini}, {Thorpe}, {Troebs}, {Vallisneri},
  {Vecchio}, {Vetrugno}, {Vitale}, {Volonteri}, {Wanner}, {Ward}, {Wass},
  {Weber}, {Ziemer}, \& {Zweifel}}]{Amaro-Seoane_2017}
{Amaro-Seoane}, P., {Audley}, H., {Babak}, S., {et~al.} 2017, arXiv e-prints,
  arXiv:1702.00786

\bibitem[{{Amaro-Seoane} {et~al.}(2007){Amaro-Seoane}, {Gair}, {Freitag},
  {Miller}, {Mandel}, {Cutler}, \& {Babak}}]{Amaro-Seoane_2007}
{Amaro-Seoane}, P., {Gair}, J.~R., {Freitag}, M., {et~al.} 2007, Classical and
  Quantum Gravity, 24, R113

\bibitem[{{Amaro-Seoane} {et~al.}(2013){Amaro-Seoane}, {Sopuerta}, \&
  {Freitag}}]{Amaro-Seoane_2013}
{Amaro-Seoane}, P., {Sopuerta}, C.~F., \& {Freitag}, M.~D. 2013, \mnras, 429,
  3155

\bibitem[{{Arcodia} {et~al.}(2021){Arcodia}, {Merloni}, {Nandra}, {Buchner},
  {Salvato}, {Pasham}, {Remillard}, {Comparat}, {Lamer}, {Ponti}, {Malyali},
  {Wolf}, {Arzoumanian}, {Bogensberger}, {Buckley}, {Gendreau}, {Gromadzki},
  {Kara}, {Krumpe}, {Markwardt}, {Ramos-Ceja}, {Rau}, {Schramm}, \&
  {Schwope}}]{Arcodia_2021}
{Arcodia}, R., {Merloni}, A., {Nandra}, K., {et~al.} 2021, \nat, 592, 704

\bibitem[{{Bardeen} {et~al.}(1972){Bardeen}, {Press}, \&
  {Teukolsky}}]{Bardeen_1972}
{Bardeen}, J.~M., {Press}, W.~H., \& {Teukolsky}, S.~A. 1972, \apj, 178, 347

\bibitem[{{Bloecker}(1995)}]{Bloecker1995}
{Bloecker}, T. 1995, \aap, 297, 727

\bibitem[{{Chen} {et~al.}(2018){Chen}, {Fishbach}, \& {Holz}}]{Chen_2018}
{Chen}, H.-Y., {Fishbach}, M., \& {Holz}, D.~E. 2018, \nat, 562, 545

\bibitem[{{Chen} {et~al.}(2021){Chen}, {Qiu}, {Li}, \& {Liu}}]{Chen2021}
{Chen}, X., {Qiu}, Y., {Li}, S., \& {Liu}, F.~K. 2021, arXiv e-prints,
  arXiv:2112.03408

\bibitem[{{Czerny} {et~al.}(2009){Czerny}, {Siemiginowska}, {Janiuk},
  {Nikiel-Wroczy{\'n}ski}, \& {Stawarz}}]{Czerny_2009}
{Czerny}, B., {Siemiginowska}, A., {Janiuk}, A., {Nikiel-Wroczy{\'n}ski}, B.,
  \& {Stawarz}, {\L}. 2009, \apj, 698, 840

\bibitem[{{Dai} \& {Blandford}(2013)}]{Dai_2013}
{Dai}, L. \& {Blandford}, R. 2013, \mnras, 434, 2948

\bibitem[{{Danzmann}(2000)}]{Danzmann_2000}
{Danzmann}, K. 2000, Advances in Space Research, 25, 1129

\bibitem[{{Dufour} {et~al.}(2017){Dufour}, {Blouin}, {Coutu},
  {Fortin-Archambault}, {Thibeault}, {Bergeron}, \& {Fontaine}}]{Dufour2017}
{Dufour}, P., {Blouin}, S., {Coutu}, S., {et~al.} 2017, in Astronomical Society
  of the Pacific Conference Series, Vol. 509, 20th European White Dwarf
  Workshop, ed. P.~E. {Tremblay}, B.~{Gaensicke}, \& T.~{Marsh}, 3

\bibitem[{{Gair} {et~al.}(2004){Gair}, {Barack}, {Creighton}, {Cutler},
  {Larson}, {Phinney}, \& {Vallisneri}}]{Gair_2004}
{Gair}, J.~R., {Barack}, L., {Creighton}, T., {et~al.} 2004, Classical and
  Quantum Gravity, 21, S1595

\bibitem[{{Gezari}(2021)}]{Gezari_2021}
{Gezari}, S. 2021, \araa, 59, 21

\bibitem[{{Giustini} {et~al.}(2020){Giustini}, {Miniutti}, \&
  {Saxton}}]{Giustini_2020}
{Giustini}, M., {Miniutti}, G., \& {Saxton}, R.~D. 2020, \aap, 636, L2

\bibitem[{{Grzedzielski} {et~al.}(2017){Grzedzielski}, {Janiuk}, {Czerny}, \&
  {Wu}}]{Grzedzielski_2017}
{Grzedzielski}, M., {Janiuk}, A., {Czerny}, B., \& {Wu}, Q. 2017, \aap, 603,
  A110

\bibitem[{{Guillochon} \& {Ramirez-Ruiz}(2013)}]{Guillochon2013}
{Guillochon}, J. \& {Ramirez-Ruiz}, E. 2013, \apj, 767, 25

\bibitem[{{Hameury} {et~al.}(1994){Hameury}, {King}, {Lasota}, \&
  {Auvergne}}]{Hameury_1994}
{Hameury}, J.~M., {King}, A.~R., {Lasota}, J.~P., \& {Auvergne}, M. 1994, \aap,
  292, 404

\bibitem[{{Herwig} {et~al.}(1999){Herwig}, {Bl{\"o}cker}, {Langer}, \&
  {Driebe}}]{Herwig1999}
{Herwig}, F., {Bl{\"o}cker}, T., {Langer}, N., \& {Driebe}, T. 1999, \aap, 349,
  L5

\bibitem[{{Hiltner} \& {Schild}(1966)}]{Hiltner_1966}
{Hiltner}, W.~A. \& {Schild}, R.~E. 1966, \apj, 143, 770

\bibitem[{{Ingram} {et~al.}(2021){Ingram}, {Motta}, {Aigrain}, \&
  {Karastergiou}}]{Ingram_2021}
{Ingram}, A., {Motta}, S.~E., {Aigrain}, S., \& {Karastergiou}, A. 2021,
  \mnras, 503, 1703

\bibitem[{{Janiuk} \& {Czerny}(2011)}]{Janiuk_2011}
{Janiuk}, A. \& {Czerny}, B. 2011, \mnras, 414, 2186

\bibitem[{{Janiuk} {et~al.}(2002){Janiuk}, {Czerny}, \&
  {Siemiginowska}}]{Janiuk_2002}
{Janiuk}, A., {Czerny}, B., \& {Siemiginowska}, A. 2002, \apj, 576, 908

\bibitem[{{Jefremov} {et~al.}(2015){Jefremov}, {Tsupko}, \&
  {Bisnovatyi-Kogan}}]{Jefremov_2015}
{Jefremov}, P.~I., {Tsupko}, O.~Y., \& {Bisnovatyi-Kogan}, G.~S. 2015, \prd,
  91, 124030

\bibitem[{{King}(2020)}]{King_2020}
{King}, A. 2020, \mnras, 493, L120

\bibitem[{{Kolb} \& {Ritter}(1990)}]{Kolb1990}
{Kolb}, U. \& {Ritter}, H. 1990, \aap, 236, 385

\bibitem[{{Kuerban} {et~al.}(2020){Kuerban}, {Huang}, {Geng}, \&
  {Zong}}]{Kuerban2020}
{Kuerban}, A., {Huang}, Y.-F., {Geng}, J.-J., \& {Zong}, H.-S. 2020, arXiv
  e-prints, arXiv:2012.05748

\bibitem[{{Law-Smith} {et~al.}(2017){Law-Smith}, {MacLeod}, {Guillochon},
  {Macias}, \& {Ramirez-Ruiz}}]{Law-Smith_2017}
{Law-Smith}, J., {MacLeod}, M., {Guillochon}, J., {Macias}, P., \&
  {Ramirez-Ruiz}, E. 2017, \apj, 841, 132

\bibitem[{{Linial} \& {Sari}(2017)}]{Linial2017}
{Linial}, I. \& {Sari}, R. 2017, \mnras, 469, 2441

\bibitem[{{Luo} {et~al.}(2016){Luo}, {Chen}, {Duan}, {Gong}, {Hu}, {Ji}, {Liu},
  {Mei}, {Milyukov}, {Sazhin}, {Shao}, {Toth}, {Tu}, {Wang}, {Wang}, {Yeh},
  {Zhan}, {Zhang}, {Zharov}, \& {Zhou}}]{Luo_2016}
{Luo}, J., {Chen}, L.-S., {Duan}, H.-Z., {et~al.} 2016, Classical and Quantum
  Gravity, 33, 035010

\bibitem[{Maggiore(2018)}]{Maggiore:2018sht}
Maggiore, M. 2018, {Gravitational Waves. Vol. 2: Astrophysics and Cosmology}
  (Oxford University Press)

\bibitem[{{Merloni} \& {Nayakshin}(2006)}]{Merloni_2006}
{Merloni}, A. \& {Nayakshin}, S. 2006, \mnras, 372, 728

\bibitem[{{Metzger} {et~al.}(2022){Metzger}, {Stone}, \&
  {Gilbaum}}]{Metzger_2021}
{Metzger}, B.~D., {Stone}, N.~C., \& {Gilbaum}, S. 2022, \apj, 926, 101

\bibitem[{{Miniutti} {et~al.}(2019){Miniutti}, {Saxton}, {Giustini},
  {Alexander}, {Fender}, {Heywood}, {Monageng}, {Coriat}, {Tzioumis}, {Read},
  {Knigge}, {Gandhi}, {Pretorius}, \& {Ag{\'\i}s-Gonz{\'a}lez}}]{Miniutti_2019}
{Miniutti}, G., {Saxton}, R.~D., {Giustini}, M., {et~al.} 2019, \nat, 573, 381

\bibitem[{{Paczynski} \& {Sienkiewicz}(1981)}]{Paczynski_1981}
{Paczynski}, B. \& {Sienkiewicz}, R. 1981, \apjl, 248, L27

\bibitem[{{Palmese} \& {Kim}(2021)}]{Palmese2021}
{Palmese}, A. \& {Kim}, A.~G. 2021, \prd, 103, 103507

\bibitem[{{Patterson} {et~al.}(2000){Patterson}, {Walker}, {Kemp},
  {O'Donoghue}, {Bos}, \& {Stubbings}}]{Patterson_2000}
{Patterson}, J., {Walker}, S., {Kemp}, J., {et~al.} 2000, \pasp, 112, 625

\bibitem[{{Paxton} {et~al.}(2011){Paxton}, {Bildsten}, {Dotter}, {Herwig},
  {Lesaffre}, \& {Timmes}}]{Paxton_2011}
{Paxton}, B., {Bildsten}, L., {Dotter}, A., {et~al.} 2011, \apjs, 192, 3

\bibitem[{{Paxton} {et~al.}(2013){Paxton}, {Cantiello}, {Arras}, {Bildsten},
  {Brown}, {Dotter}, {Mankovich}, {Montgomery}, {Stello}, {Timmes}, \&
  {Townsend}}]{Paxton_2013}
{Paxton}, B., {Cantiello}, M., {Arras}, P., {et~al.} 2013, \apjs, 208, 4

\bibitem[{{Paxton} {et~al.}(2015){Paxton}, {Marchant}, {Schwab}, {Bauer},
  {Bildsten}, {Cantiello}, {Dessart}, {Farmer}, {Hu}, {Langer}, {Townsend},
  {Townsley}, \& {Timmes}}]{Paxton_2015}
{Paxton}, B., {Marchant}, P., {Schwab}, J., {et~al.} 2015, \apjs, 220, 15

\bibitem[{{Paxton} {et~al.}(2018){Paxton}, {Schwab}, {Bauer}, {Bildsten},
  {Blinnikov}, {Duffell}, {Farmer}, {Goldberg}, {Marchant}, {Sorokina},
  {Thoul}, {Townsend}, \& {Timmes}}]{Paxton_2018}
{Paxton}, B., {Schwab}, J., {Bauer}, E.~B., {et~al.} 2018, \apjs, 234, 34

\bibitem[{{Peters}(1964)}]{Peters_1964}
{Peters}, P.~C. 1964, Physical Review, 136, 1224

\bibitem[{{Reines} \& {Volonteri}(2015)}]{Reines_2015}
{Reines}, A.~E. \& {Volonteri}, M. 2015, \apj, 813, 82

\bibitem[{{Ritter}(1988)}]{Ritter1988}
{Ritter}, H. 1988, \aap, 202, 93

\bibitem[{{Rosswog} {et~al.}(2009){Rosswog}, {Ramirez-Ruiz}, \&
  {Hix}}]{Rosswog_2009}
{Rosswog}, S., {Ramirez-Ruiz}, E., \& {Hix}, W.~R. 2009, \apj, 695, 404

\bibitem[{{Sander} {et~al.}(2012){Sander}, {Hamann}, \& {Todt}}]{Sander_2012}
{Sander}, A., {Hamann}, W.~R., \& {Todt}, H. 2012, \aap, 540, A144

\bibitem[{{Sari} \& {Fragione}(2019)}]{Sari_2019}
{Sari}, R. \& {Fragione}, G. 2019, \apj, 885, 24

\bibitem[{{Sepinsky} {et~al.}(2007){Sepinsky}, {Willems}, \&
  {Kalogera}}]{Sepinsky_2007}
{Sepinsky}, J.~F., {Willems}, B., \& {Kalogera}, V. 2007, \apj, 660, 1624

\bibitem[{{Shen}(2019)}]{Shen_2019}
{Shen}, R.-F. 2019, \apjl, 871, L17

\bibitem[{{Sheng} {et~al.}(2021){Sheng}, {Wang}, {Ferland}, {Shu}, {Yang},
  {Jiang}, \& {Chen}}]{Sheng_2021}
{Sheng}, Z., {Wang}, T., {Ferland}, G., {et~al.} 2021, \apjl, 920, L25

\bibitem[{{Shu} {et~al.}(2018){Shu}, {Wang}, {Dou}, {Jiang}, {Wang}, \&
  {Wang}}]{Shu_2018}
{Shu}, X.~W., {Wang}, S.~S., {Dou}, L.~M., {et~al.} 2018, \apjl, 857, L16

\bibitem[{{Shu} {et~al.}(2017){Shu}, {Wang}, {Jiang}, {Wang}, {Sun}, \&
  {Zhou}}]{Shu_2017}
{Shu}, X.~W., {Wang}, T.~G., {Jiang}, N., {et~al.} 2017, \apj, 837, 3

\bibitem[{{Sigurdsson} \& {Rees}(1997)}]{Sigurdsson_1997}
{Sigurdsson}, S. \& {Rees}, M.~J. 1997, \mnras, 284, 318

\bibitem[{{Thorne}(1974)}]{Thorne_1974}
{Thorne}, K.~S. 1974, \apj, 191, 507

\bibitem[{{Wang} \& {Merritt}(2004)}]{Wang_2004}
{Wang}, J. \& {Merritt}, D. 2004, \apj, 600, 149

\bibitem[{{Wang} {et~al.}(2018){Wang}, {Wang}, \& {Zou}}]{Wang_2018}
{Wang}, Y.~Y., {Wang}, F.~Y., \& {Zou}, Y.~C. 2018, \prd, 98, 063503

\bibitem[{{Wang} {et~al.}(2019){Wang}, {Wang}, {Zou}, \& {Dai}}]{Wang_2019}
{Wang}, Y.~Y., {Wang}, F.~Y., {Zou}, Y.~C., \& {Dai}, Z.~G. 2019, \apjl, 886,
  L22

\bibitem[{{Wu} {et~al.}(2016){Wu}, {Czerny}, {Grzedzielski}, {Janiuk}, {Gu},
  {Dong}, {Cao}, {You}, {Yan}, \& {Sun}}]{Wu_2016}
{Wu}, Q., {Czerny}, B., {Grzedzielski}, M., {et~al.} 2016, \apj, 833, 79

\bibitem[{{Xiao} {et~al.}(2011){Xiao}, {Barth}, {Greene}, {Ho}, {Bentz},
  {Ludwig}, \& {Jiang}}]{Xiao_2011}
{Xiao}, T., {Barth}, A.~J., {Greene}, J.~E., {et~al.} 2011, \apj, 739, 28

\bibitem[{{Yu} {et~al.}(2018){Yu}, {Ratra}, \& {Wang}}]{Yu2018}
{Yu}, H., {Ratra}, B., \& {Wang}, F.-Y. 2018, \apj, 856, 3

\bibitem[{{Zalamea} {et~al.}(2010){Zalamea}, {Menou}, \&
  {Beloborodov}}]{Zalamea_2010}
{Zalamea}, I., {Menou}, K., \& {Beloborodov}, A.~M. 2010, \mnras, 409, L25

\end{thebibliography}
\end{document}